\documentclass[prl,aps,showpacs,twocolumn,amsmath,amssymb]{revtex4}
\usepackage{epsfig}
\usepackage{graphicx}
\usepackage{amssymb,amsmath}
\usepackage{subfigure}
\usepackage{bbm}
\usepackage[dvips]{color}

\newcommand{\surfacePlotScale}[0]{0.23}

\begin{document}

\title{Quantum Entangled Dark Solitons Formed by Ultracold Atoms in Optical Lattices}
\author{R. V. Mishmash$^{1,2}$ and L. D. Carr$^{2}$}
\affiliation{$^1$Department of Physics, University of California, Santa Barbara, CA 93106, USA \\
$^2$Department of Physics, Colorado School of Mines, Golden, CO
80401, USA}
\date{\today}

\begin{abstract}
Inspired by experiments on Bose-Einstein condensates in optical lattices, we
study the quantum evolution of dark soliton initial conditions in
the context of the Bose-Hubbard Hamiltonian. An extensive set of quantum
measures is utilized in our analysis, including von Neumann and generalized
quantum entropies, quantum depletion, and the pair correlation function. We find that quantum effects cause the soliton to fill in. Moreover, soliton-soliton collisions become inelastic, in strong contrast to the predictions of mean-field theory.  These features show that the lifetime and collision properties of dark solitons in optical lattices provide clear signals of quantum effects.
\end{abstract}


\maketitle

Systems of ultracold atoms loaded into optical lattices offer an excellent experimental and theoretical test bed for the study of complex quantum many-body phenomena, including out-of-equilibrium quantum dynamics. The unprecedented tunability of system parameters such as lattice height, filling, dimensionality, and species allows one to experimentally simulate key condensed matter Hamiltonians~\cite{Jaksch05_AnnPhys_315_52}. This control also permits study of \emph{dynamical} properties of the system, an aspect of quantum lattice physics not typically accessible in the solid state. Examples include the experiment of the nonadiabatic transition across the Mott-superfluid border~\cite{Greiner02_Nature_415_39}, the experiment of strongly damped quantum transport in a combined optical lattice and harmonic trap~\cite{Fertig05_PRL_94_120403}, and theoretical work investigating relaxation properties of the system after a quantum quench \cite{Kollath07_PRL_98_180601}.

In this Letter, we present a quantum many-body study of the dynamics of dark solitons formed by ultracold bosonic gases in one-dimensional (1D) optical lattices.  Solitons are localized, persistent, nonlinear waves which appear throughout nature.  Dark solitons formed by Bose-Einstein condensates (BECs) were first observed in 1999 \cite{Burger99_PRL_83_5198} and have lately been the topic of exciting experimental research \cite{Sengstock08_NaturePhys_4_496, Sengstock08_PRL_101_120406, Oberthaler08_PRL_101_130401}.  Owing to negligible quantum depletion out of the condensed mode, BECs in the mean-field limit are well-described by the Gross-Pitaevskii (GP) or nonlinear Schr\"{o}dinger equation (NLS).  Such a mean-field description has proven to be an excellent model for describing ground state properties \cite{Dalfovo99_RevModPhys_71_463} and soliton dynamics \cite{Sengstock08_PRL_101_120406, Anderson01_PRL_86_2926, Feder00_PRA_62_053606} for experiments in harmonic traps.  However, finite temperature \cite{Jackson07_PRA_75_051601} and quantum fluctuations \cite{Dziarmaga03_JPhysB_36_1217} have both been shown to affect the stability of dark solitons in such continuous geometries.

In the presence of an optical lattice, one can still attempt to describe BECs within a mean-field picture, either by directly employing the continuous NLS with an external lattice potential \cite{Bronski01_PRL_89_1402, Efremidis03_PRA_67_063608} or by applying a lowest Bloch band tight-binding approximation to the condensate order parameter in the continuous GP equation \cite{Trombettoni01_PRL_86_2353}.  The latter procedure results in the discrete nonlinear Schr\"{o}dinger equation (DNLS).  On the other hand, the Bose-Hubbard Hamiltonian (BHH) \cite{Fisher89_PRB_40_546} is a discretization of the full many-body Hamiltonian that can be almost perfectly realized with a system of ultracold bosons in an optical lattice \cite{Jaksch98_PRL_81_3108}.  In fact, the DNLS is most perspicuously obtained as a mean-field approximation of the BHH \cite{Amico98_PRL_80_2189}.  In this work, we consider quantum entangled dynamical evolution of the dark soliton---a robust, emergent property of the corresponding mean-field theory (DNLS)---by performing quasiexact simulations of the BHH using newly available quantum algorithms \cite{Vidal04_PRL_93_040502}.  Hence, we directly address how many-body effects such as quantum fluctuations and quantum entanglement affect the stability of dark solitons, thereby providing a \emph{quantitative} measure of the applicability of mean-field theory in describing such dynamics.

The Bose-Hubbard Hamiltonian describing ultracold bosons in a 1D optical lattice reads
\begin{equation}
\hat{H} = -J \sum_{k=1}^{M-1}(\hat{b}_{k+1}^\dagger \hat{b}_{k} + \mathrm{H.c.}) + \frac{U}{2} \sum_{k=1}^{M} \hat{n}_k (\hat{n}_k-\hat{\mathbbm{1}}), \label{eqn:bhh}
\end{equation}
where $J$ is the hopping strength and $U$ is the on-site atom-atom interaction energy.  Equation (\ref{eqn:bhh}) assumes box boundary conditions on a lattice of $M$ sites.  The bosonic destruction and creation operators $\hat{b}_k$ and $\hat{b}_k^\dagger$ obey the usual bosonic commutation relations, and $\hat{n}_k\equiv \hat{b}_k^\dagger\hat{b}_k$ is the number operator that counts the number of bosons at site $k$.  The (quasi-)1D regime is reached experimentally by making a 3D lattice and ramping up the lattice heights in the transverse directions creating an \emph{array of 1D tubes} with a sinusoidal potential in the longitudinal direction.  Typically, each tube contains 10-100 atoms and each site in the tube only one or a few atoms.

To obtain the DNLS from the BHH, one can evolve the bosonic destruction operator $\hat{b}_k$ forward in time in the Heisenberg picture using Eq.~(\ref{eqn:bhh}):  $i\hbar\partial_t\hat{b}_k=[\hat{b}_k,\hat{H}]$.  Assuming zero quantum fluctuations obtainable via a tensor product of on-site Glauber coherent states, $\hat{b}_k$ can be replaced with its expectation value $\psi_k\equiv\langle\hat{b}_k\rangle$, resulting in the DNLS to describe the condensate order parameter:
\begin{equation}
i\hbar\partial_t \psi_k = -J\left (\psi_{k+1} + \psi_{k-1}\right) + U|\psi_k|^2 \psi_k.
\end{equation}
The solution is normalized to $N_{\mathrm{DNLS}} \equiv\sum_{k=1}^M|\psi_k|^2$.

In this work, we consider two methods for generating dark soliton initial conditions in the full many-body BHH.  In method 1, we work within a truncated, non-number-conserving Fock space that allows each site to contain at most $d-1$ bosons.  Then, via constrained imaginary time relaxation, we calculate the fundamental dark soliton solution $\{\psi_k\}$ of the DNLS and carry it over to Fock space by using a product of \emph{truncated coherent states} of the form $|\Psi\rangle = \bigotimes_{k=1}^M |z_k\rangle\mathrm{,~where~ }|z_k\rangle= \mathcal{N}_d\,\exp{(-|z_k|^2/2)}\sum_{n=0}^{d-1}\frac{z_k^n}{\sqrt{n!}}|n\rangle $; $\mathcal{N}_d$ is a normalization factor required by our truncation and $\psi_k=\langle\hat{b}_k\rangle\approx z_k$.  In method 2, we generalize the methods of density and phase manipulation for soliton creation \cite{Carr01_PRA_63_051601} to the BHH by first using imaginary time propagation with an external Gaussian potential to dig a density notch and then applying an instantaneous phase gradient across the notch \cite{Mishmash09_long}.  For simulation of the BHH in real and imaginary time, we use Vidal's time-evolving block decimation (TEBD) algorithm \cite{Vidal04_PRL_93_040502} retaining Schmidt basis sets of size $\chi$.  This method is equivalent to a time-adaptive density matrix renormalization group routine based on matrix product states \cite{Daley04_JStatMech_P04005, White04_PRL_93_076401}.  Its accuracy depends explicitly on the amount of spatial entanglement in the states being simulated \cite{Vidal03_PRL_91_147902}.

To characterize the quantum nature of the system we employ six distinct measures.  (i) The \emph{quantum depletion} describes the occupation of non-condensate modes. The natural orbitals of the system are the eigenvectors of the one-body density matrix $\langle\hat{b}^\dagger_j \hat{b}_i\rangle$, where we denote the $k$th component of the $(j+1)$th most highly occupied natural orbital as $\phi^{(j)}_k$ with corresponding occupation $N_j$.  The condensate wave function is the eigenvector $\phi^{(0)}$ of the one-body density matrix whose occupation $N_0$ is largest \cite{Penrose56_PR_104_576}. Depletion out of the condensate mode is defined as $D\equiv 1-N_0/N_{\mathrm{avg}}$, where $N_{\mathrm{avg}}\equiv\sum_{k=1}^M\langle \hat{n}_k\rangle$ is the total average number.  (ii) The \emph{order parameter} $\langle\hat{b}_k\rangle$ maps directly onto the DNLS dependent variable for infinite-dimensional coherent states and has a time-dependent norm $N_b\equiv\sum_{k=1}^M |\langle\hat{b}_k\rangle|^2$ that measures the system coherence.  In the thermodynamic limit in 1D, it is well known that the local order parameter $\langle\hat{b}_k\rangle$ vanishes; however, Bose condensation is possible in finite quasi-1D systems, and products of coherent states with $\langle\hat{b}_k\rangle\neq 0$ serve as a mean-field description of such condensed states.  (iii) The \emph{on-site expected particle number} $\langle\hat{n}_k\rangle$ is the local density actually measured in experiment. (iv) The \emph{local von Neumann entropy}, or entropy of entanglement, is an entanglement measure defined as $S_{\mathrm{vN},k}\equiv-\mathrm{Tr} \left[ \hat{\rho}_k\log_d(\hat{\rho}_k) \right ] \in [0, 1]$ with $\hat{\rho}_k\equiv \mathrm{Tr}_{j\neq k}\hat{\rho}$. It measures the entanglement of the $k$th localized mode with the rest of the lattice. (v) The \emph{average local impurity} ~\cite{Brennen03_QuantInfoAndComp_3_619, Barnum04_PRL_92_107902} $Q\equiv \frac{d}{d-1} \left[1 - \frac{1}{M}\sum_{k=1}^M \mathrm{Tr}(\hat{\rho}_k^2) \right] \in [0, 1]$ describes how far the full system is from a local pure state at each site. (vi) Finally, the \emph{pair correlation function} $g^{(2)}_{ij}\equiv\langle\hat{b}_i^\dagger\hat{b}_j^\dagger\hat{b}_j\hat{b}_i\rangle $ measures the joint probability of measuring two particles at sites $i$ and $j$.

\begin{figure}[b]
\begin{center}
\hspace{-0.04in}
\subfigure{\scalebox{\surfacePlotScale}{\includegraphics[angle=0]{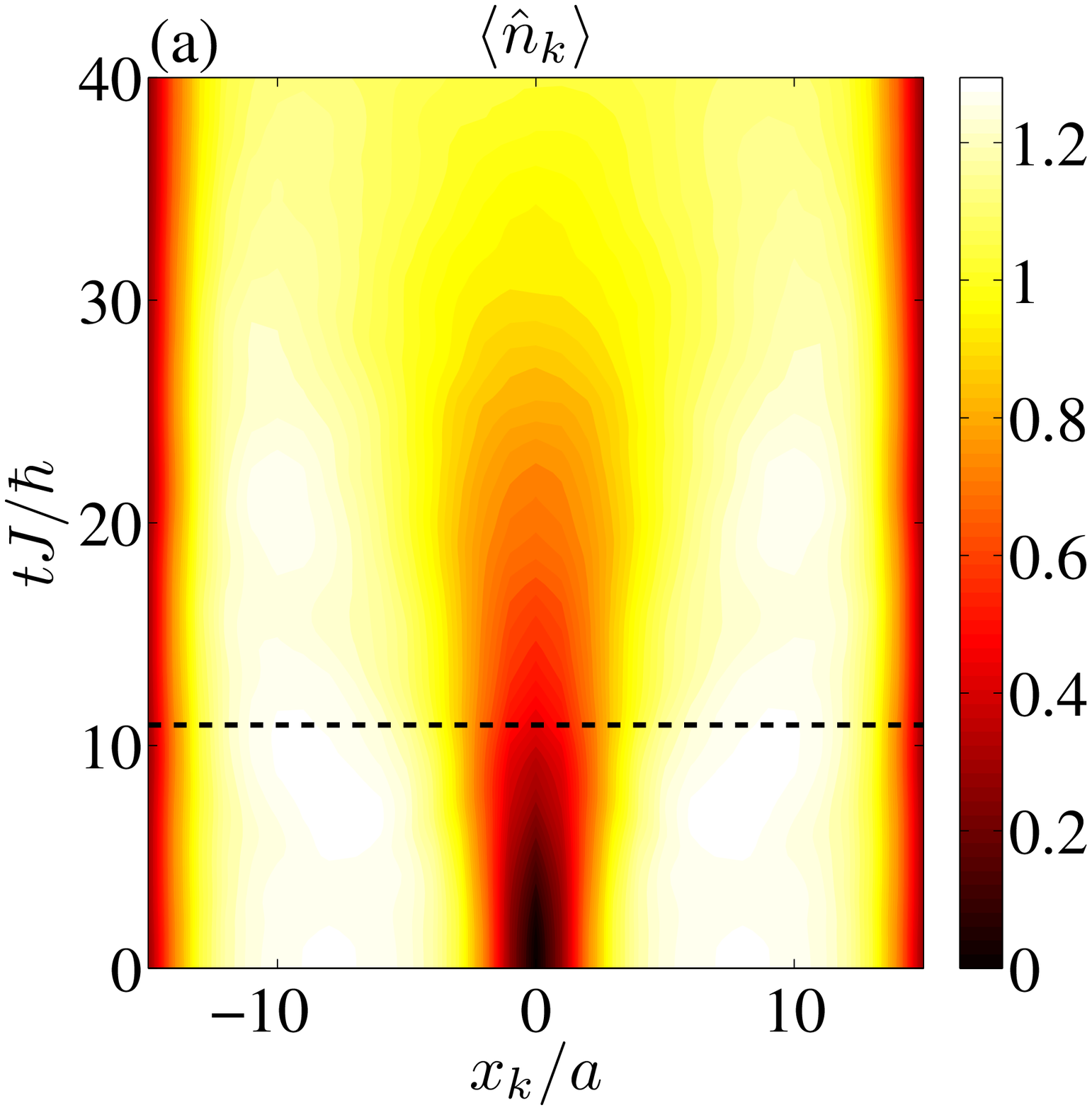}}}
\hspace{-0.05in}
\subfigure{\scalebox{\surfacePlotScale}{\includegraphics[angle=0]{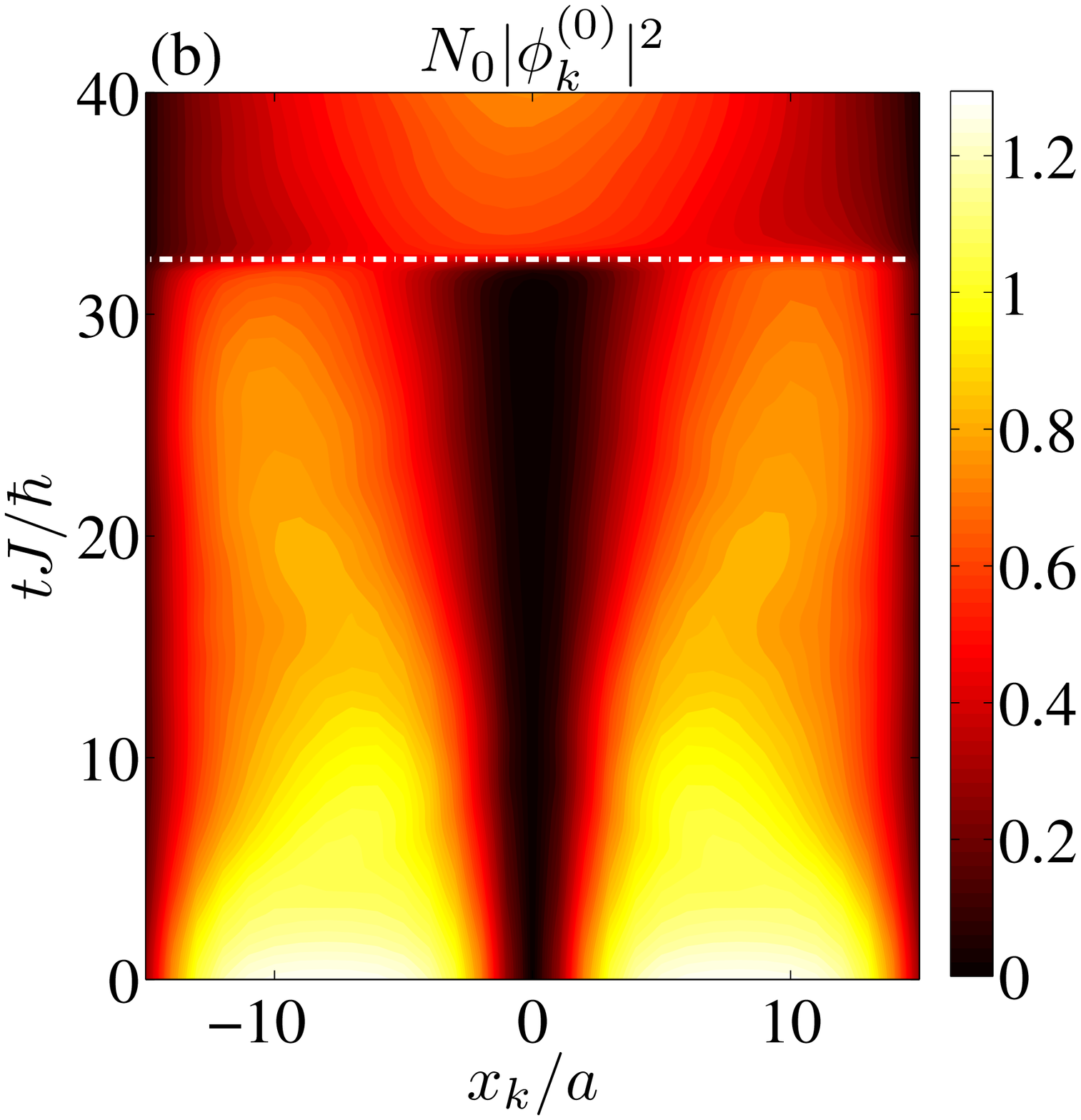}}}
\end{center}
\vspace{-0.30in}
\begin{center}
\subfigure{\scalebox{\surfacePlotScale}{\includegraphics[angle=0]{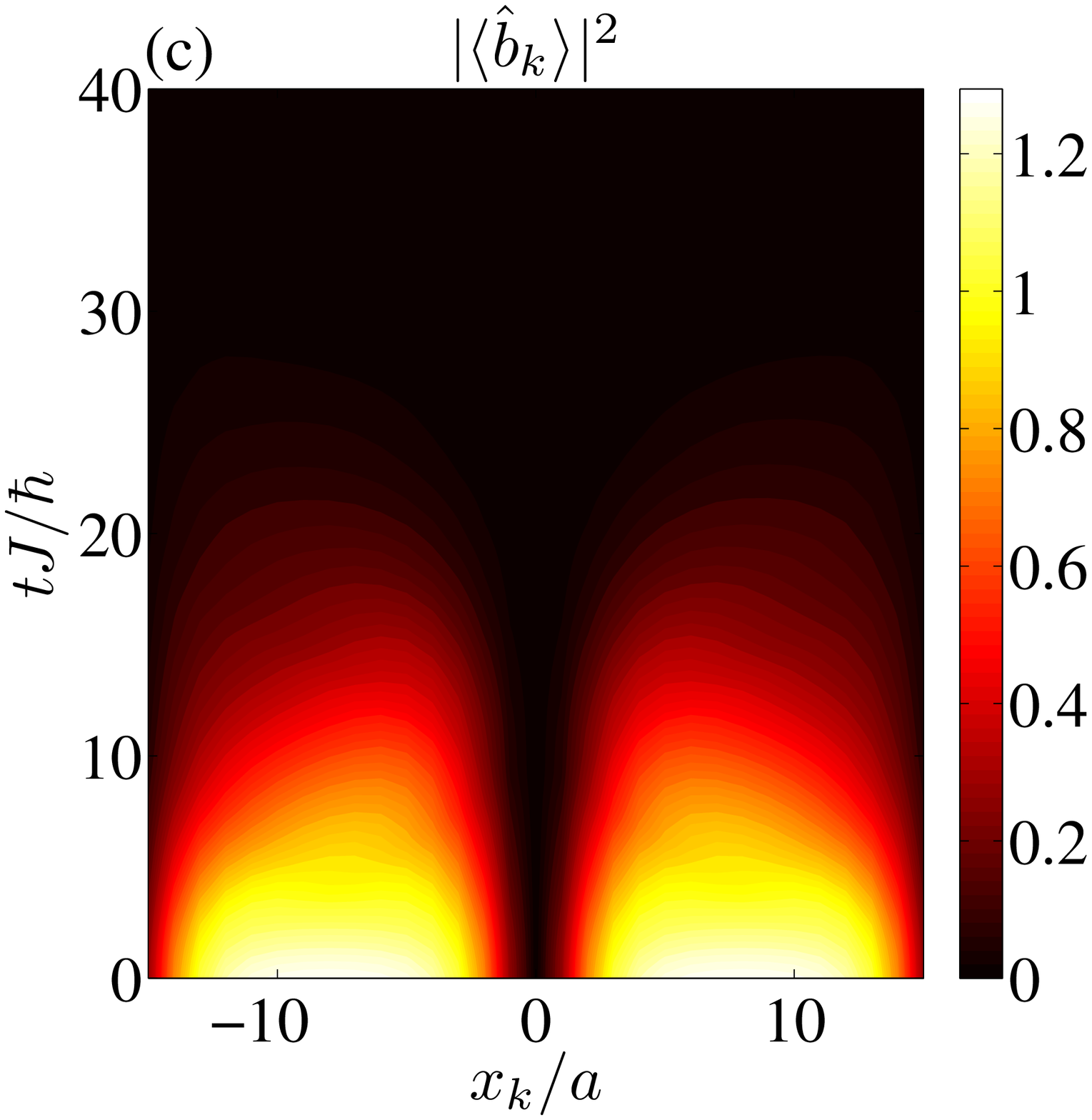}}}
\hspace{-0.07in}
\subfigure{\scalebox{0.23}{\includegraphics[angle=0]{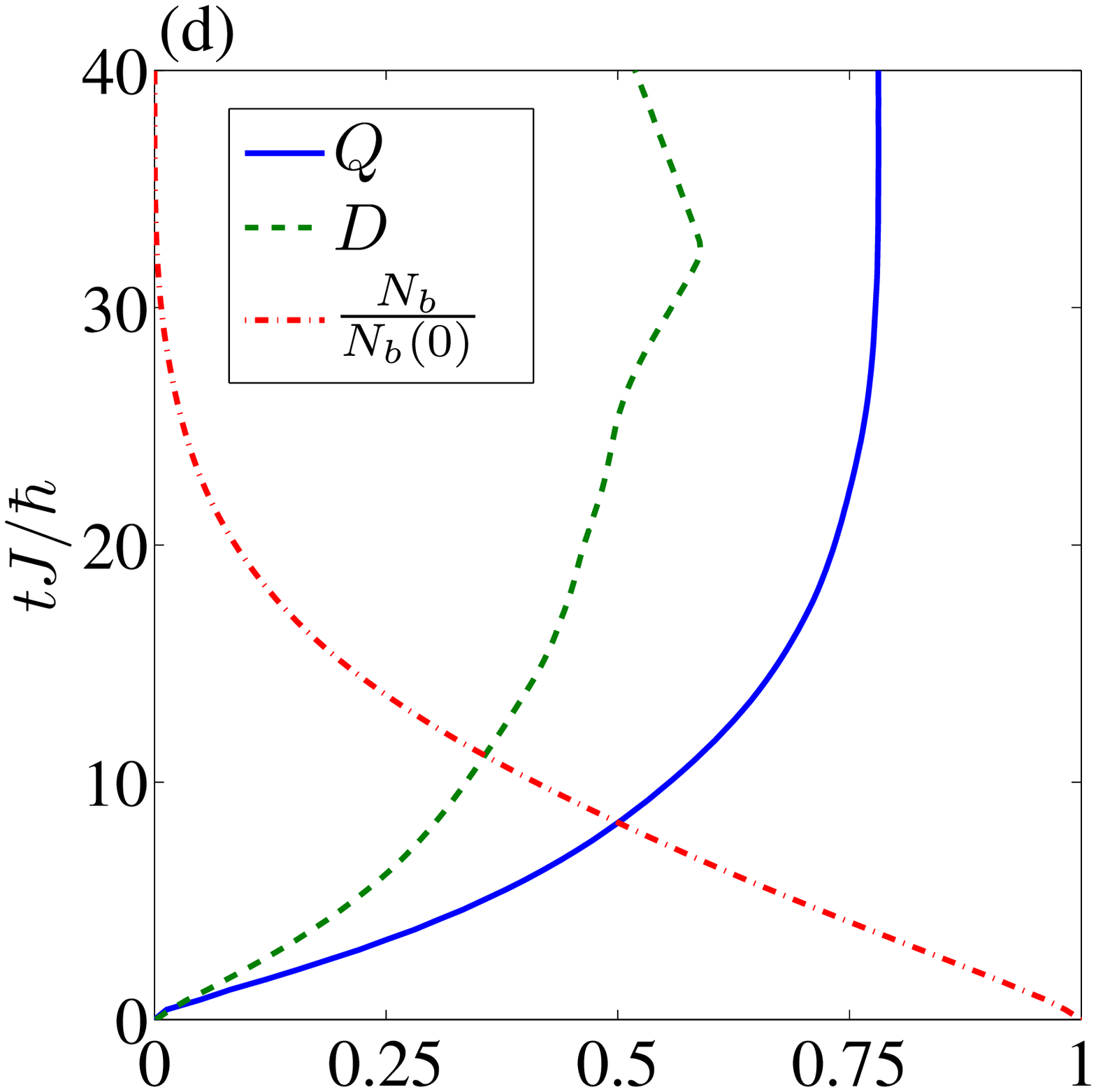}}}
\end{center}
\vspace{-0.25in}
\caption{(color online) \emph{Three ways to describe a quantum
soliton:} (a) particle number density, (b) Penrose-Onsager condensate wave
function density, and (c) order parameter density versus position and time
during quantum evolution of a dark soliton initial state obtained via method 1.  Shown in (d) is the time dependence of the average local impurity, quantum depletion, and order parameter norm. \label{fig:standing1}} \vspace{-0.05in}
\end{figure}

Figure \ref{fig:standing1} depicts a characteristic simulation of standing quantum soliton propagation for initial conditions obtained via method 1 as described above.  The parameters are $\nu U/J=0.35$ at filling $\nu\equiv N_{\mathrm{DNLS}}/M=1$ for $M=31$ lattice sites with $\chi=50$ and $d=7$.  For all results, we have checked for sufficient convergence in all numerical parameters, i.e., $\chi$, $d$, as well as $\delta t$, the Trotter time step size used in the TEBD \cite{Vidal04_PRL_93_040502}.  We refer to the parameter $\nu U/J$ as the \emph{effective interaction strength} because it accounts for both the atom density $\nu$ and the interaction energy $U$, all scaled to $J$. At the initial time when the system is in a product of coherent states resembling a standing soliton, the depletion as calculated by diagonalizing the one-body density matrix is negligible ($D\approx 0.1\%$). The condensate wave function according to the Penrose-Onsager definition closely resembles the stationary DNLS soliton solution, and this mode is occupied by all but $DN_\mathrm{avg}$ bosons. However, unitary evolution according to the BHH causes an increase in occupation of non-solitonic orbitals giving the soliton a finite lifetime.  The second most highly occupied natural orbital $\phi^{(1)}$ is a density \emph{maximum} that fills in the soliton notch \cite{Mishmash09_long}.

These quantum dynamics can be explained by noting that the standing dark soliton is an antisymmetric wave function, and there are allowed two-body scattering processes which preserve the symmetry of the many-body wave function that can deplete particles out of the soliton mode.  For instance, two particles initially in the dark soliton orbital can scatter into a symmetric orbital which need not necessarily vanish at the lattice center.  We see in Fig. \ref{fig:standing1} that such processes are energetically preferable over time as particles deplete into the mode $\phi^{(1)}$ as well as orbitals of higher order.  That is, our theory also treats higher order processes such as two atoms previously depleted into a symmetric mode scattering into an antisymmetric mode of higher order than the soliton mode \cite{Mishmash09_long}.  At time $tJ/\hbar\approx 33$, $\phi^{(1)}$ gains higher occupation than the soliton mode initially denoted $\phi^{(0)}$.  This crossing in occupation numbers of natural orbitals is indicated by the discontinuity and corresponding dash-dotted white line in Fig. \ref{fig:standing1}(b).  The decay of the order parameter density shown in Fig. \ref{fig:standing1}(c) closely follows the collapse of the soliton structure depicted in Fig. \ref{fig:standing1}(a) where the black dashed line corresponds to the exponential decay time $\tau_b$ of the order parameter norm $N_b$.  In contrast, the evolution of the condensate order parameter according to the DNLS (not shown) reveals a stably propagating density notch, and the initial DNLS soliton solution is dynamically stable \cite{Mishmash09_long}.  All in all, the behavior described above is general, and the qualitative picture remains when a harmonic trap potential is added.

\begin{figure}[b]
\vspace{-0.05in}
\begin{center}
\hspace{-0.0in}\includegraphics[scale=0.25,angle=0]{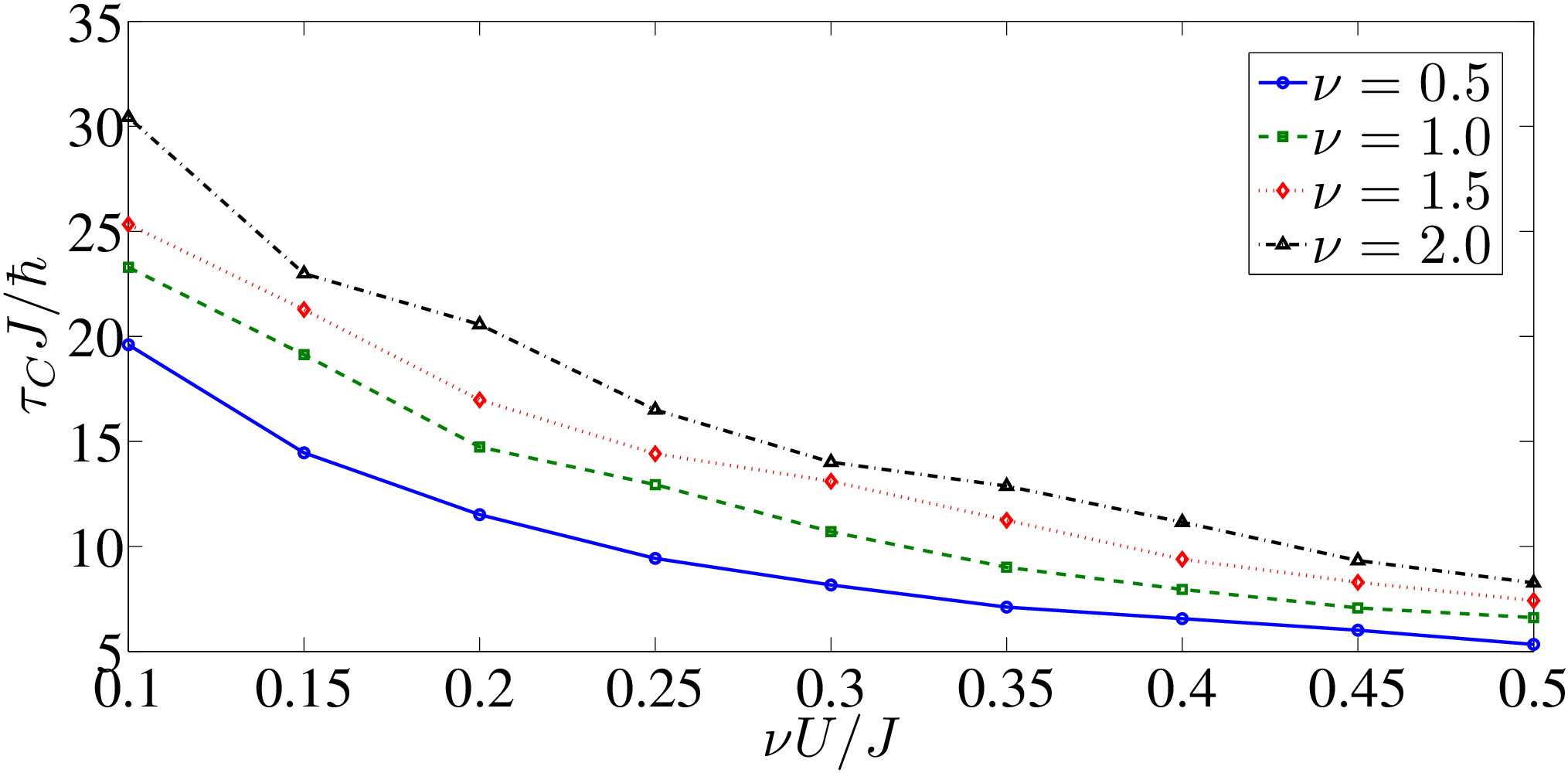}
\end{center}
\vspace{-0.2in}\caption{(color online) \emph{Quantum soliton lifetimes.} The decay time $\tau_c$ of the soliton contrast versus interaction strength $\nu U/J$ at four separate filling factors $\nu$.  Here, we use data corresponding to dark solitons created via method 2 with $M=30$ lattice sites \cite{Mishmash09_long}.  All values of $\nu U/J$ considered reside within the superfluid region of the BHH ground state
phase diagram.  \label{fig:cohtimes}} \vspace{-0.125in}
\end{figure}

In Fig. \ref{fig:cohtimes}, we depict the decay times $\tau_C$ of the soliton contrast for a range of $\nu U/J$ values at four filling factors: $\nu=0.5,1.0,1.5,~\mathrm{and}~2.0$.  The soliton contrast at a given time is defined as $C\equiv(\langle \hat{n}_\mathrm{max}\rangle-\langle \hat{n}_\mathrm{mid}\rangle)/(\langle \hat{n}_\mathrm{max}\rangle+\langle \hat{n}_\mathrm{mid}\rangle)$, where $\langle \hat{n}_\mathrm{max}\rangle$ is the maximum of the average density over all sites $k$ and $\langle \hat{n}_\mathrm{mid}\rangle$ is the density at the lattice center.  The time scale $\tau_C$ is the time at which $C$ decays to a value of 1/2.  In order to achieve higher filling factors, we use method 2 of soliton generation with a number-conserving TEBD routine.  Convergence of these results requires taking up to $\chi=120$ and $d=9$.  It is clear that the soliton lifetime decreases with increased interaction strength and increases with increased filling factor.

\begin{figure}[b]
\begin{center}
\subfigure{\scalebox{\surfacePlotScale}{\includegraphics[angle=0]{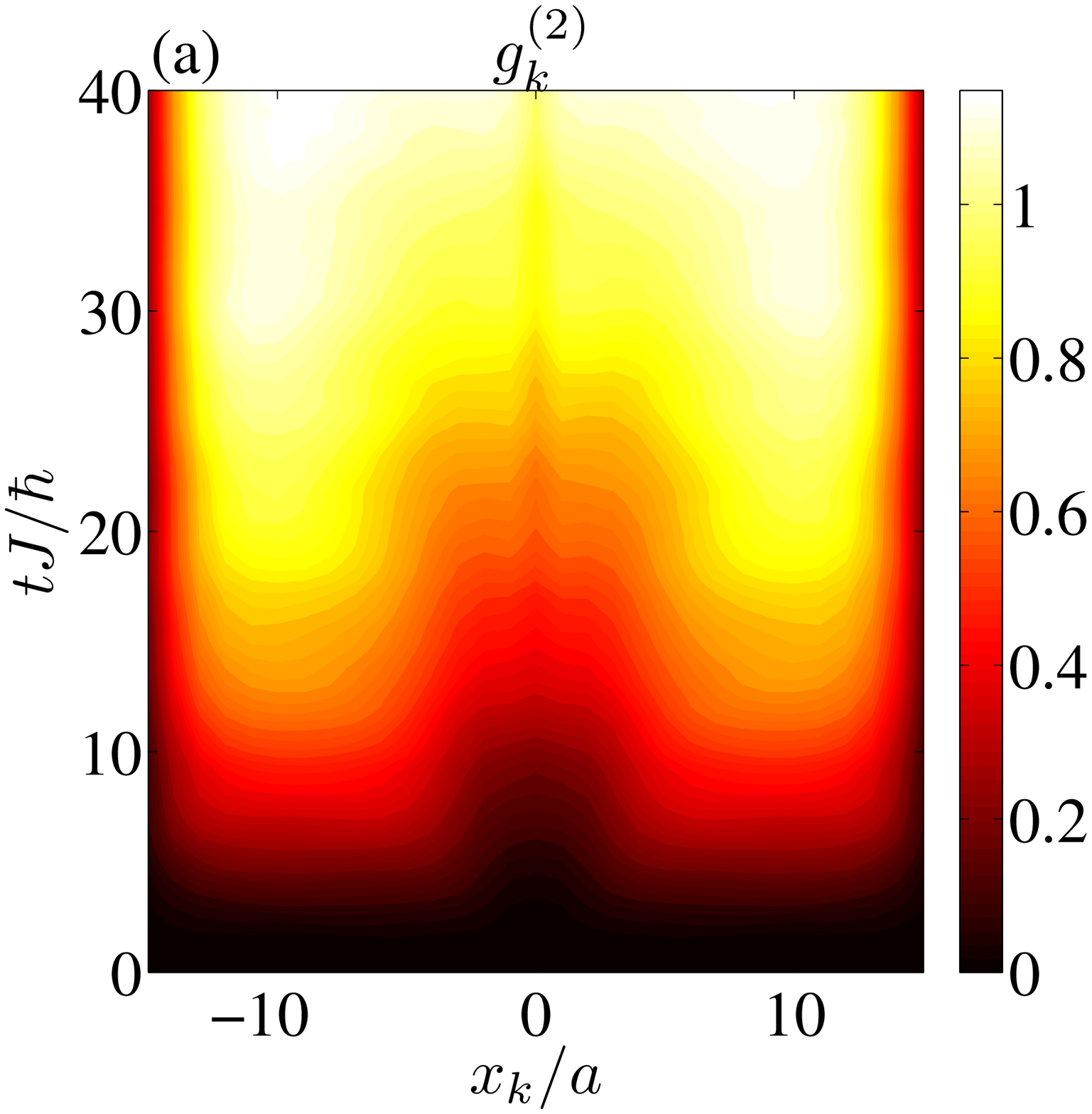}}}
\hspace{-0.05in}
\subfigure{\scalebox{\surfacePlotScale}{\includegraphics[angle=0]{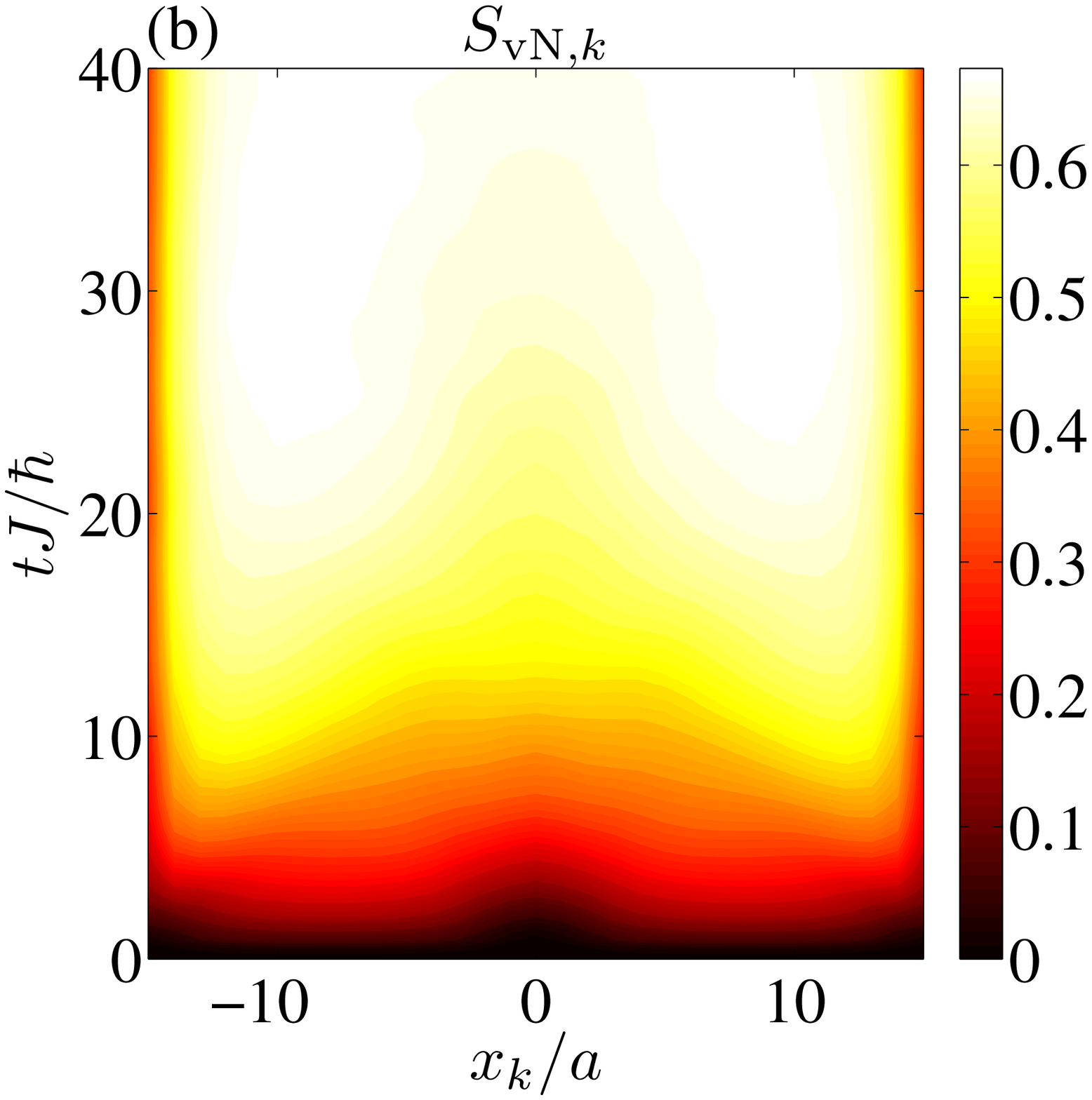}}}
\end{center}
\vspace{-0.25in}
\caption{(color online) \emph{Many-body characterization of a quantum soliton:  pair correlations and entanglement entropy.} The (a) pair correlation function and (b) Local von Neumann entropy for the same simulation presented in Fig. \ref{fig:standing1}.  \label{fig:standing2}}
\end{figure}

Our approach allows a full quantum many-body characterization of the dynamics. In Fig. \ref{fig:standing1}(d), we see that even though the average local impurity and quantum depletion vanish initially due to the nature of the initial state, these measures grow over time.  This is also true for the local von Neumann entropy as shown in Fig. \ref{fig:standing2}(b).  Finally, in Fig. \ref{fig:standing2}(a) we calculate $g_k^{(2)}\equiv g^{(2)}_{\mathrm{mid},\,k}$, the pair correlation function of site $k$ with the center lattice site, i.e., the initial position of the soliton.  This measure does not remain zero over time at all sites and is largest in the decaying soliton's tails; thus, the soliton decay phenomenon observed above is not the result of averaging a wandering, yet nondecaying, dark soliton over many measurements.

\begin{figure}[t]
\begin{center}
\subfigure{\scalebox{\surfacePlotScale}{\includegraphics[angle=0]{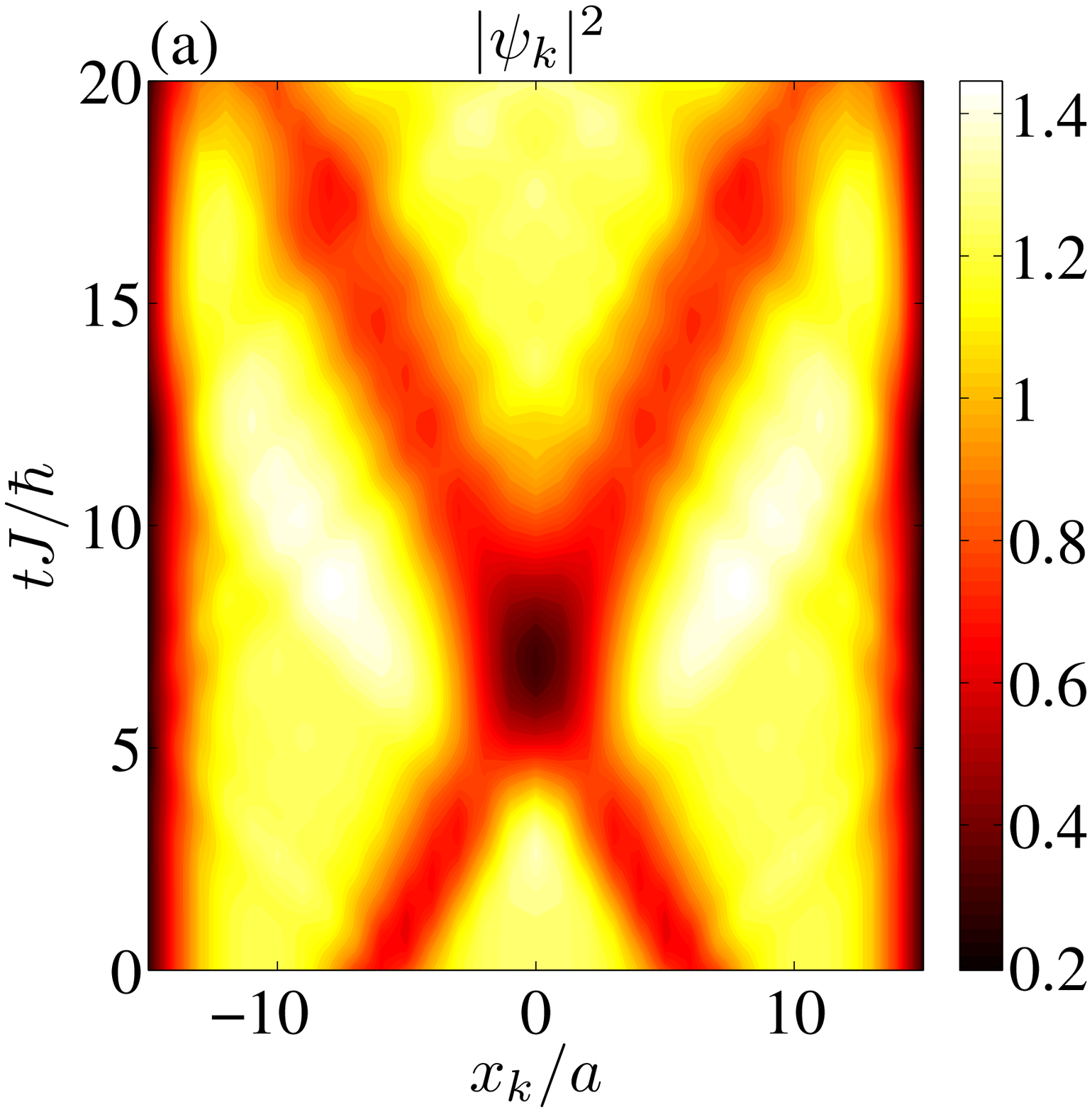}}}
\hspace{-0.05in}
\subfigure{\scalebox{\surfacePlotScale}{\includegraphics[angle=0]{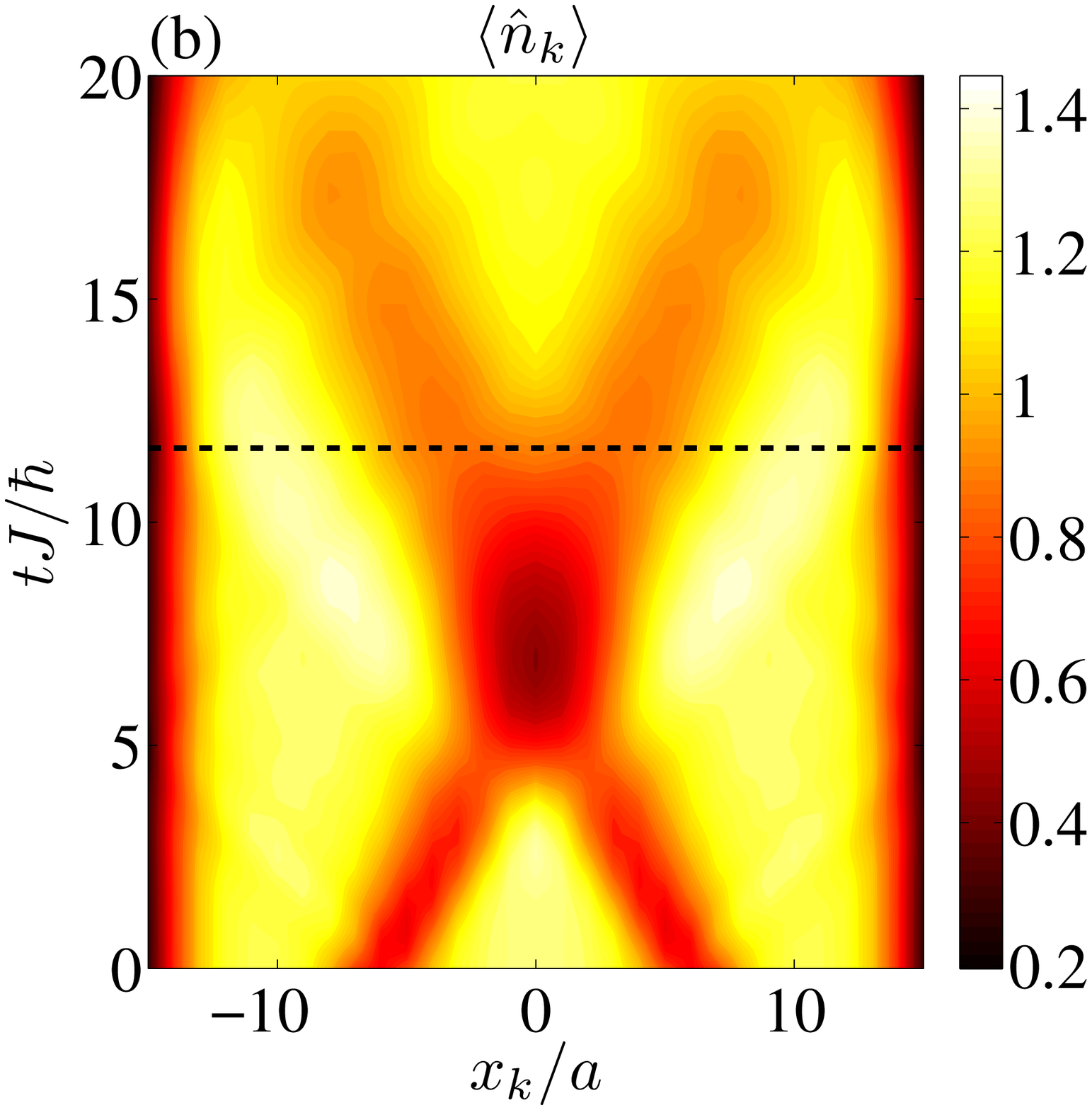}}}
\end{center}
\vspace{-0.30in}
\begin{center}
\subfigure{\scalebox{\surfacePlotScale}{\includegraphics[angle=0]{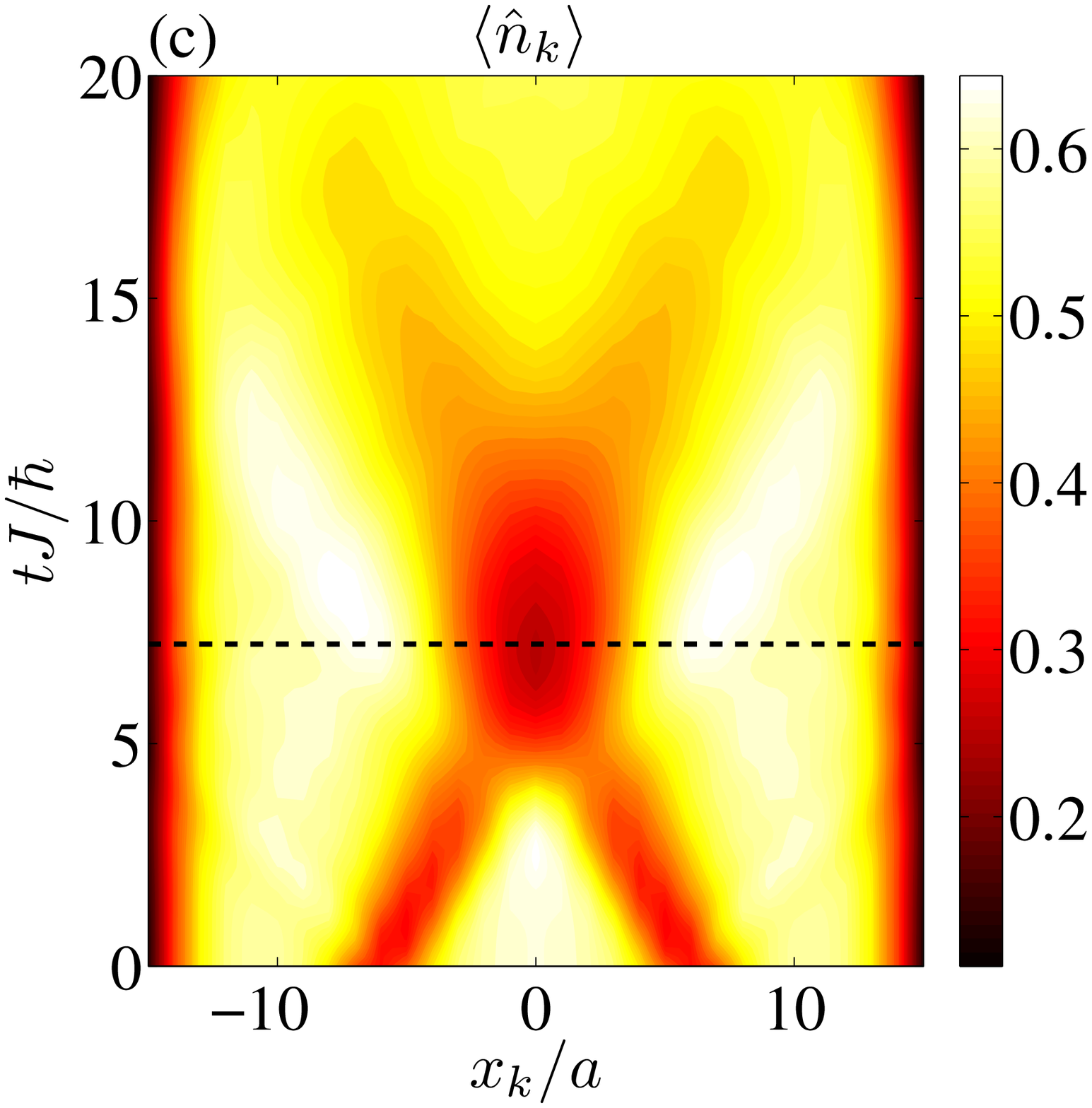}}}
\hspace{-0.05in}
\subfigure{\scalebox{\surfacePlotScale}{\includegraphics[angle=0]{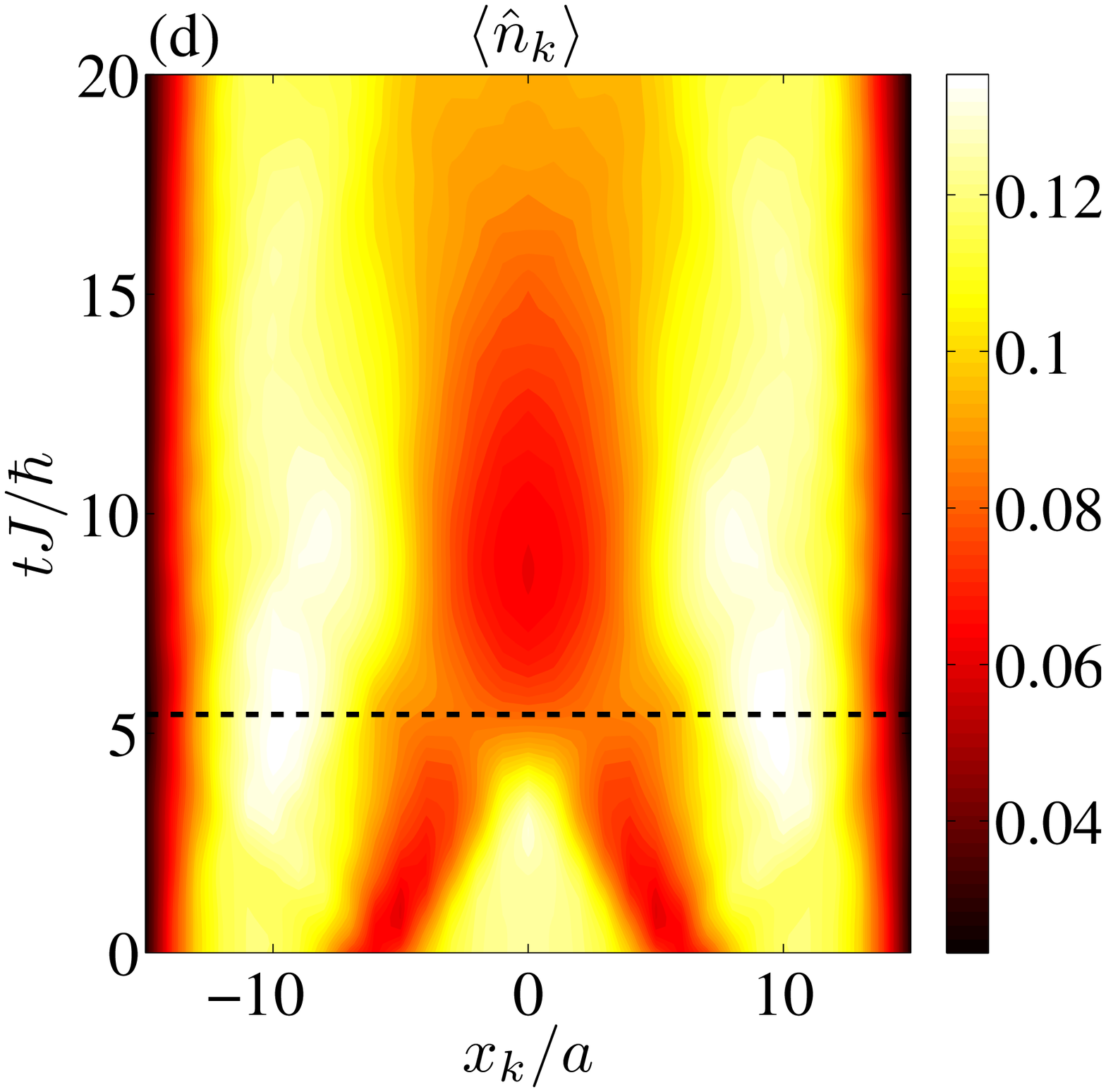}}}
\end{center}
\vspace{-0.25in} \caption{(color online) \emph{Quantum-induced inelasticity.}  Dark soliton collision for (a) the DNLS and the corresponding quantum evolution at filling factors (b) $\nu=1$, (c) $\nu=0.5$, and (d) $\nu=0.1$ at effective interaction strength $\nu U/J=0.35$ for $M=31$ sites.  For (b), (c), and (d), the decoherence time $\tau_b$ [black dashed line, cf. Fig. \ref{fig:standing1}(a)] is tuned to occur after, during, and before the collision time, respectively. \label{fig:colliding1}}
\end{figure}

In Fig. \ref{fig:colliding1} we illustrate soliton-soliton collisions as described by both the DNLS and the BHH. The initial conditions are obtained by applying the methods of density and phase engineering to the DNLS using Gaussian potentials for the density engineering and hyperbolic tangent phase profiles for the phase engineering and then carrying the state to Fock space via truncated coherent states, as in method 1 for standing solitons.  Figure \ref{fig:colliding1}(a) depicts the DNLS dynamics of a near-elastic soliton collision, and Fig. \ref{fig:colliding1}(b) shows the density during the corresponding quantum evolution.  In this case, the collision occurs before the decoherence time $\tau_b$ and the average particle number closely follows the mean-field order parameter density:  the collision is still very elastic in accordance with mean-field theory.  In the DNLS, the filling $\nu$ only changes the norm of the solution for fixed $\nu U/J$; however, in the quantum picture, we can vary the time of decoherence by varying the filling as evident in Fig. \ref{fig:cohtimes}. We show in Fig. \ref{fig:colliding1}(c)-(d) that the elasticity of a collision of two quantum solitons \emph{decreases} when the decoherence time becomes comparable to the collision time. That is, the solitons interact or ``stick together'' for a longer time. In these cases, the second most dominant natural orbital $\phi^{(1)}$ is a standing soliton, so that increased occupation of this mode has the effect of increasing the time over which the solitons collide.  Subsequent filling in of the notch after collision can be explained by bosons occupying even higher order modes, i.e., $\phi^{(2)}$, $\phi^{(3)}$, etc.  The fact that the solitons can be made to stick together is another compelling piece of evidence that we are not observing diffusing solitons averaged over many measurements.

The increase in occupation of higher order natural orbitals, i.e., depletion out of the condensate wave function, is a completely many-body effect that cannot be accounted for using an unperturbed mean-field theory such as the DNLS.  However, the lowest-energy mode of the Bogoliubov spectrum for a stationary dark soliton solution of the DNLS can be an anomalous mode with a density maximum in the center of the lattice \cite{Dziarmaga03_JPhysB_36_1217}. This Bogoliubov mode resembles, but is not equivalent to, the second order mode $\phi^{(1)}$ observed for the standing soliton case \cite{Mishmash09_long}.  By simulating the BHH, a true quantum field theory, we are able to calculate explicitly the time dependence of the distribution of the natural orbitals as
well as their exact spatial form.

By calculating the stability times of dark solitons, which are stable structures in the DNLS but nonequilibrium states of the BHH, we are evaluating the validity of using the DNLS to describe the system dynamics.  It is thought that in the superfluid regime of the Bose-Hubbard phase diagram mean-field theory should be applicable.  However, we show that there is always a time at which quantum fluctuations cause such a model to break down, providing specific predictions in Fig. \ref{fig:cohtimes}.  As one tunes toward the Mott border, the stability times decrease, indicating an even stronger presence of quantum fluctuations and a further breakdown of mean-field theory.  As $U/J$ is increased past the region studied in Fig.~\ref{fig:cohtimes}, one eventually obtains a soliton that extends only over a single site, necessitating a
multi-band BHH.

In conclusion, we have constructed quantum many-body analogs of dark solitons and analyzed their dynamics. This was achieved by using Vidal's TEBD algorithm to simulate real time evolution in the Bose-Hubbard Hamiltonian of initial dark solitonlike many-body states.  We showed that quantum effects cause a finite soliton lifetime and induce an inelasticity in soliton-soliton collisions.

We thank Charles Clark, Ippei Danshita, Michael Wall, and Jamie Williams for useful discussions.  This material is based upon work supported by the National Science Foundation under Grant No. PHY-0547845, as part of the NSF CAREER program.  RVM acknowledges support from Microsoft Station Q and the SURF program at NIST.

%\bibliographystyle{prsty}
%\bibliography{../../References/00Database}

\end{document}